\documentclass[journal]{IEEEtran}
\usepackage{cite}
\usepackage{etoolbox}
\usepackage{mathtools, cuted}
\usepackage{tabu}
\usepackage{array}
\usepackage{comment}
\usepackage{flushend}

\ifCLASSINFOpdf
\else
\fi

\usepackage{amsmath, amsthm, amssymb, amsfonts}
\usepackage{xcolor}

\usepackage{graphicx}
\usepackage{epstopdf}

\makeatletter
\DeclareRobustCommand\bfseriesitshape{%
\not@math@alphabet\itshapebfseries\relax
\fontseries\bfdefault
\fontshape\itdefault
\selectfont
}
\makeatother

\DeclareTextFontCommand{\textbfit}{\bfseriesitshape}

\begin{document}

\title{ Asymptotic Performance Analysis of Generalized User Selection for Interference-Limited Multiuser Secondary Networks}

\author{\IEEEauthorblockN{ Yazan H. Al-Badarneh, Costas N. Georghiades, Mohamed-Slim Alouini}
%\IEEEauthorblockA{%$^{1}$ Department of Electrical and Computer Engineering \\
%Texas A\&M University, College Station, TX 77843\\
%Emails:\{albadarneh, georghiades, carlos.mejia\}@tamu.edu}
}
\maketitle

\begin{abstract}
We analyze the asymptotic performance of a generalized multiuser diversity scheme for an interference-limited secondary multiuser network of underlay cognitive radio systems. Assuming a large number of secondary users and that the noise at each secondary user's receiver is negligible compared to the interference from the primary transmitter, the secondary transmitter transmits information to the $k$-th best secondary user, namely, the one with the $k$-th highest signal-to-interference ratio (SIR). We use extreme value theory to show that the $k$-th highest SIR converges uniformly in distribution to an inverse gamma random variable for a fixed $k$ and large number of secondary users. We use this result to derive asymptotic expressions for the average throughput, effective throughput, average bit error rate and outage probability of the $k$-th best secondary user under continuous power adaptation at the secondary transmitter, which ensures satisfaction of the instantaneous interference constraint at the primary receiver caused by the secondary transmitter. Numerical simulations show that our derived asymptotic expressions are accurate for different values of system parameters.

%We analyze the asymptotic performance of a multiuser diversity scheme for an interference-limited secondary multiuser network of underlay cognitive radio systems. Assuming a large number of secondary users and that the noise at each secondary user's receiver is negligible compared to the interference from the primary transmitter, the secondary transmitter transmits information to the $k$-th best secondary user, namely, the one with the $k$-th highest signal-to-interference ratio. We use extreme value theory to derive the asymptotic distribution of the $k$-th highest signal-to-interference ratio for a fixed $k$ and large number of secondary users. We use this result to analyze the asymptotic average throughput, effective throughput, average bit error rate and outage probability for the $k$-th best secondary user under continuous power adaptation at the secondary transmitter, which ensures satisfaction of the instantaneous interference constraint at the primary receiver caused by the secondary transmitter.
\end{abstract}

\begin{IEEEkeywords}
Extreme value distributions; Fading channels; Cognitive radio; Multiuser diversity; Communication system performance. 
\end{IEEEkeywords}

\section{Introduction}
Cognitive radio (CR) is an important technology to maximize radio spectrum utilization efficiency\cite{819467}\nocite{4840529}-\cite{1391031}. In CR systems, the secondary network is allowed to share the spectrum allocated to the primary network provided that the interference caused by the secondary transmitter (ST) does not deteriorate the performance of the primary network. Consequently, the challenge is to maintain the interference at the primary receiver (PR) below a pre-determined threshold level. This can be achieved by adapting the ST transmit power that ensures satisfaction of the interference constraint at the PR \cite{4786456}. 

Multiuser diversity is considered an important diversity technique to improve wireless communication systems performance \cite{tse2005fundamentals}. Considering a multiuser network where the users experience independent fading conditions, the basic idea of multiuser diversity is to select the users with the best fading conditions for transmission or reception to obtain a specific performance gain. Multiuser diversity in CR systems has attracted much attention recently. Researchers analyze the performance of multiuser diversity techniques for uplink multiuser underlay CR systems without taking the interference from the primary network into consideration\cite{4786488}\nocite{ekin2012capacity}\nocite{li2013capacity} \nocite{khan2015performance}\nocite{7881835}-\cite{AGHAZADEH2018160}. In particular, the ergodic capacity (throughput) of multiuser diversity gain of uplink multiuser underlay CR systems is investigated in \cite{4786488}. In \cite{ekin2012capacity}, the authors analyze the achievable capacity gain of uplink multiuser spectrum-sharing systems over dynamic fading environments. In \cite{li2013capacity}, the outage probability and effective capacity are analyzed for opportunistic spectrum sharing in Rayleigh fading environment. In \cite{khan2015performance}, the authors analyze the outage probability, average symbol error rate (SER) and ergodic capacity of an opportunistic multiuser cognitive network with multiple primary users assuming the channels in the secondary network are independent but not identical Nakagami-$m$ fading. In \cite{7881835}, \cite{AGHAZADEH2018160} the authors analyze the outage probability and average capacity of multiuser diversity in single-input multiple-output (SIMO) spectrum sharing systems. The mentioned previous works did not consider the impact of interference from the primary network to the secondary network. However, in the practical CR systems the interference from the primary network will greatly affect the performance of the secondary network. The ergodic capacity of multiuser diversity in CR systems with interference from the primary network is investigated in \cite{5403611}. Recently, the ergodic capacity of various multiuser scheduling schemes in downlink cognitive radio networks with interference from the primary network is analyzed in \cite{sibomana2016ergodic}; here, the authors analyze the ergodic capacity of multiuser diversity scheduling under the outage constraint of multiple primary user receivers and the secondary user (SU) maximum transmit power limit. 

 When the interference from the primary transmitter (PT) is much larger than the noise at the secondary receiver, the performance of the secondary network is limited by the interference from the primary transmitter and the quantity of interest is signal-to-interference ratio (SIR). Such CR system can be described as interference-limited underlay CR system \cite{6924725}. For the interference-limited underlay CR systems considered in \cite{6924725}, the authors analyze the average bit error rate (BER) and outage probability for receive antenna selection schemes under discrete power adaptation at the ST.

%  Numerical simulations show that the derived asymptotic expressions accurate even for not so large (realistic) number of secondary users. This suggests that asymptotic analysis presented in this paper is  very useful to real systems and practical situations.} %We validate the accuracy of the asymptotic expressions through numerical simulation.} 

\subsection{Motivation} 
The mentioned previous works have only focused on conventional multiuser diversity in underlay CR systems where the SU with the best link quality is selected. Furthermore, no prior work has considered the performance of the conventional multiuser diversity for interference-limited underlay CR systems with continuous power adaptation at the ST. Accordingly, we focus in this paper on a generalized multiuser diversity scheme that features selection of the $k$-th best SU for interference-limited secondary multiuser network under continuous power adaptation. The $k$-th best SU selection is of practical interest in underlay CR systems since the best SU may not be selected under given traffic conditions.  This might happen when the best user is unavailable or occupied by other service requirements \cite{LI2015745}, in handoff situations \cite{6146496} or due to scheduling delay \cite{6171803}. Clearly, the $k$-th best SU selection scheme includes the best SU selection at $k=1$ as a special case.

 In general, it is difficult to analyze the exact performance of the $k$-th best SU selection scheme for arbitrary number of secondary users. Hence, we use the extreme value theorem (EVT) \cite{david2003order} to analyze the asymptotic performance (in the limit of large number of secondary users) of such selection scheme. As we will show later, EVT provides tractable and accurate  asymptotic expressions for the average throughput, effective throughput, average bit error rate and outage probability. The derived asymptotic expressions are accurate for practical CR systems with not so large (realistic) number of secondary users. We validate the accuracy of the asymptotic expressions through numerical simulation.

%Numerical simulations show that the derived asymptotic expressions accurate even for not so large (realistic) number of secondary users.  This suggests that our asymptotic analysis is  very useful to real systems and practical situations.}

It should be noted that the derived expressions are obtained by assuming perfect channel state information (CSI) of the secondary transmitter to primary receiver channel. The impact of imperfect CSI on the mathematical analysis is also investigated. As we will discuss later, the derived mathematical expressions under the perfect CSI can be used to deduce the system performance under the imperfect CSI case.

\subsection{Related work and new contributions} 
In our previous work \cite{8269400}, we used EVT to derive simple closed-form asymptotic expressions for the average throughput, effective throughput and average BER for the link with the $k$-th highest signal-to-noise ratio (SNR) in traditional wireless communication systems with no spectrum sharing. We showed that the $k$-th highest SNR converges uniformly in distribution to a $\textit{Log-Gamma}$ random variable \footnote{ If $X$ is   a Gamma random variable, then we say that $Y= \log(X)$ is a $\textit{Log-Gamma}$ random variable whose support is the real line \cite{leemis2008univariate}.}. As a special case, if $k=1$, the $\textit{Log-Gamma}$ reduces to the Gumbel random variable. The average throughput, effective throughput and average BER were derived for various channel models that are widely used to characterize fading in wireless communication systems such as Weibull, Gamma, $\alpha -\mu$ and Gamma-Gamma.

Unlike\cite{8269400}, in this paper we consider a downlink interference-limited secondary multiuser network, where the noise at each secondary user receiver is negligible compared to the interference from the PT. We assume that the ST transmits information to the $k$-th best secondary user (SU), namely, the SU with the $k$-th highest SIR. Meanwhile, the ST adopts continuous limited power adaptation strategy to satisfy the instantaneous interference constraint at the PR. Our contribution is to utilize EVT to analyze the performance of the $k$-th best SU for underlay CR systems. More specifically, we show that  the $k$-th highest SIR converges uniformly in distribution to an inverse gamma random variable for a fixed $k$ and large number of secondary users\footnote{Here, the $k$-th highest SIR converges to an inverse gamma random variable. This is different from \cite{8269400}, where the $k$-th highest SNR converges to a $\textit{Log-Gamma}$ random variable.}. Then, we derive novel closed-form asymptotic expressions for the average and effective throughputs of the $k$-th best SU employing continuous power adaptation at the ST with both limited and unlimited transmit power \footnote{ A portion of this work was accepted for publication  in 2018 IEEE Vehicular Technology Conference (VTC) \cite{yazanVTC2018}. In \cite{yazanVTC2018}, we only studied the asymptotic average and effective throughputs of the $k$-th best secondary user selection in uplink multiuser cognitive radio systems under unlimited transmit power. The VTC paper is available online at https://arxiv.org/pdf/1804.05257v2.pdf.  Different from \cite{yazanVTC2018}, in this paper we analyze the average and effective throughputs, average BER and outage probability of the $k$-th best secondary user selection for interference limited CR systems under both limited and unlimited ST transmit power adaptation strategies. We also investigate the impact of imperfect CSI of the secondary transmitter to primary receiver channel.  }.  Furthermore, novel closed-form asymptotic expressions for the average BER and outage probability with continuous power adaptation and unlimited ST power are derived.  It should be noted that although we use EVT approach in this paper, the analysis and the derived expressions are completely different from what were obtained for traditional wireless communication systems with no spectrum sharing.

% It should be noted that we adopt EVT in this work due to the difficulty in obtaining closed- form tractable expressions for the average and effective throughputs, average BER and outage probability for the $k$-th best SU selection scheme. EVT provides tractable and very accurate expressions for different system parameters. The derived expressions are obtained by assuming perfect channel state information (CSI) of the secondary transmitter to primary receiver channel. The impact of imperfect CSI on the mathematical analysis is also investigated. As we will show in later, the derived mathematical expressions under the perfect CSI can be used to deduce the system performance under the imperfect CSI case.

The rest of this paper is organized as follows. In Section II we discuss the system model. In Section III we analyze the asymptotic average throughput, effective throughput, average BER and  outage probability of the $k$-th best SU. Section IV includes numerical results and Section V concludes.

\section{System Model}
 As shown in Fig. 1, we consider an underlay secondary network consisting of one ST equipped with a single antenna, and $N$ secondary users each equipped with a single antenna. The secondary network is sharing the spectrum of a primary network with one PT and one PR. The PT and PR are equipped with a single antenna each. Let $g_{i}$ denote the channel gain from the PT to the $i$-th secondary user's receiver (SU-Rx), where $i=1, 2, ..., N$. Let $h_{0}$ and $h_{i}$ denote the channel gain from the ST to  the PR and the $i$-th SU-Rx, respectively. We assume that the primary network is far away from the secondary network and therefore $|h_{0}|$ and $|g_{i}|$ are assumed to be independent Rayleigh distributed random variables. This implies that the channel power gains, $|h_{0}|^{2}$ and $|g_{i}|^{2}$ have  probability density functions (PDFs) $f_{0}(x) = \eta e^{-\eta x} u(x) $ and $g(x) = \lambda e^{-\lambda x}u(x)$, respectively, where $u(x)$ is the unit step function and the parameters $\eta$ and $\lambda$ are the fading parameters. The channel power gains in the secondary network, $|h_{i}|^{2}$, for $i=1, 2, ..., N$, are assumed to be independent and identically distributed (i.i.d.) Gamma random variables with PDF \footnote{ We assume that $h_{i}$ is a Nakagami-m random variable; hence, $|h_{i}|^{2}$ is a Gamma random variable. This a is a generalized fading model that includes many practical scenarios. First, Nakagami-m fading includes the Rayleigh fading as a special case when $m=1$, then  the distribution of $|h_{i}|$ becomes consistent with the distributions of $|h_{0}|$ and $|g_{i}|$. Second, in the situation where line of-sight path exists between the secondary transmitter and secondary receivers the natural choice to use the Rician distribution to model the line of-sight effect. It is well known that the Rician distribution can be accurately approximated by the Nakagami-m distribution. Motivated by these reasons, we adopt the Nakagami-m fading to model the fading in the secondary network}

\begin{figure}[h]\label{fig:1}
\begin{center}
\includegraphics[scale=0.85]{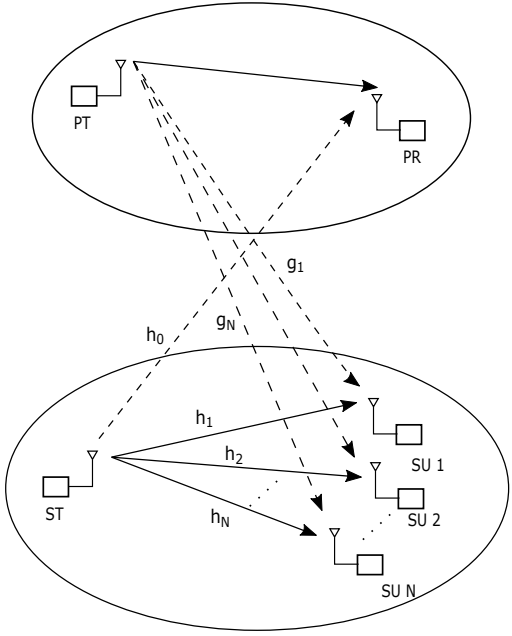}
\caption{ An underlay cognitive radio network with an ST serving $N$ secondary users.  }
\end{center}
\end{figure}

\begin{equation} \label{eq:1}
f(x)=\frac{x^{m-1}}{\beta^m \Gamma(m)} e^{-\frac{x}{\beta}} u(x). 
\end{equation}
where the parameters $m$ and $\beta$ are positive reals and $ \Gamma(m)$ is the Gamma function.

Similar to \cite{4786456}, \cite{ekin2012capacity}, \cite{khan2015performance}, \cite{6924725}, \cite{7890994} and \cite{6134707}, it is assumed that the ST has perfect CSI regarding the secondary transmitter to primary receiver channel, $h_{0}$. The ST can be informed about $h_{0}$ through a mediate band manager between PR and ST \cite{peha2005approaches} or by considering proper signaling \cite{kang2010optimal}. However, the impact of imperfect CSI on the performance of the $k$-th best SU will be discussed later in this paper. % is left for Section VI. %; thus, the derived results throughout this paper are optimistic compared to the results with imperfect CSI. 
 
%However, it is hard for the ST to obtain this CSI perfectley \cite{xu2013effective}.

With a perfect knowledge of  $|h_{0}|^{2}$, we consider a continuous power adaptation policy at the ST to control its interference to the PR such that the instantaneous transmit power of the ST is
\begin{equation} \label{eq:FFFF}
 P=\min \left(P_{S}, \frac{T}{|h_{0}|^{2}}\right),
\end{equation}
 where $P_{S}$ is the maximum instantaneous power available at the ST and $T$ is the maximum tolerable interference level at the PR. % The transmit power of the ST, $P$, is a random variable that can be modeled as either discrete or continuous random variable. 
Assuming the noise at the $i$-th SU-Rx is negligible compared to the interference from the PT, then the ST will select the $k$-th best SU; namely, the SU with the $k$-th highest SIR; i.e.,
\begin{equation} \label{eq:2}
i^{*} = \arg \text{$k$-th}\max\limits_ {i}  \left\{P Z_{i} \right\}_{i=1}^{N}
\end{equation}
where $Z_{i} =\frac{|h_{i}|^{2}}{P_{M} |g_{i}|^{2}}$, $P_{M}$ is the transmit power of the PT and $P_{M} |g_{i}|^{2}$ is the PT interference power at the $i$-th SU-Rx. 

Let $ P Z_{\left(N-k+1\right)}$ denote the instantaneous SIR at the $k$-th best  SU-Rx, where $Z_{(1)} \leq Z_{(2)} \leq .... \leq Z_{(N)}$. According to \cite{david2003order}, the PDF of $Z_{\left(N-k+1\right)}$ can be expressed in terms of the PDF, $f(z)$, and cumulative distribution function (CDF), $F(z)$, of $Z_{i}$ as 
\begin{equation}\label{eq:3}
f_{Z_{\left(N-k+1\right)}}(x)=k \binom{N}{k} f(z) F(z)^{N-k} \left(1-F(z) \right)^{k-1}, 
\end{equation}
where the CDF and PDF of $Z_{i}$ are given by \cite{6924725} 
\begin{equation}\label{eq:4}
F(z)=\left( \frac{P_{M} z}{\lambda \beta +P_{M} z}\right)^{m} u(z), 
\end{equation}
\begin{equation}\label{eq:5}
f(z)=\frac{m \lambda \beta\left(P_{M}\right)^{m} z^{m-1}}{\left( \lambda \beta +P_{M} z \right)^{m+1}} u(z), 
\end{equation}
respectively. Let $R_{(N-k+1)}=B \log_{2}(1+P Z_{\left(N-k+1\right)})$ denote the instantaneous throughput of the $k$-th best SU, where  $R_{(1)} \leq R_{(2)} \leq .... \leq R_{(N)}$ and $B$ is the system bandwidth. Then, the average throughput of the $k$-th best SU, $E\left[R_{(N-k+1)}\right]$, can be evaluated as 
\begin{equation}\label{eq:6}
E\left[{R_{(N-k+1)}} \right]=E\left[\log_{2}(1+P Z_{\left(N-k+1\right)})\right]
\end{equation}
in \text{bit/s/Hz}. The expectation in (\ref{eq:6}) is taken over the joint distribution of random variables $P$ and $ Z_{\left(N-k+1\right)}$. 
Assuming a block fading channel, the effective throughput that can be supported by a wireless system under a statistical QoS constraint described by the delay QoS exponent $\theta$ is given by \cite{1210731}
\begin{eqnarray} \label{eq:7}
\alpha (\theta) =-\frac{1}{\theta T} \log \left( E\left[ e^{-\theta T R }\right] \right),  \ \theta> 0,
\end{eqnarray}
where $R$ is a random variable which represents the instantaneous throughput during a single block and $T$ is the block length. $\theta=0$ implies there is no delay constraint and the effective throughput is then the ergodic (average) throughput of the corresponding wireless channel.  Hence, the effective throughput of the $k$-th best SU, $\alpha(\theta, k, N)$, can be expressed as %\cite{6006584}
\begin{gather} \label{eq:8}
\begin{split}
\alpha (\theta, k, N)&=-\frac{1}{A} \log_{2} \left(  E\left[ \left( 1+ P Z_{(N-k+1)} \right)^{-A} \right] \right), 
\end{split}
\end{gather}
in \text{bit/s/Hz}, where $A=\theta T/ \ln(2)$ and the expectation is taken over the joint distribution of $P$ and $Z_{(N-k+1)}$.% Applying  L'hospital's rule, one can show that $\lim_{\theta \to 0}  \alpha (\theta, k, N)= E\left[R_{(N-k+1)}\right]$, as stated before.  

If we conisder a general class of modulation schemes whose conditional BER, $P_{e}$, is given by 
\begin{gather} \label{eq:9}
\begin{split}
P_e= c\, e^{-v Y},
\end{split}
\end{gather}
where $c$ and $v$ are positive constants and $Y$ is a random variable which represents the instantaneous received SIR, the average BER of the $k$-th best SU can be expressed as
\begin{gather} \label{eq:10}
\begin{split}
\overline{ P_{e}}(k,N)=c\, E\left[ e^{-v PZ_{(N-k+1)}}\right], %C \mathcal{M} _{X_{(N-k+1)}} (-g \rho), 
\end{split}
\end{gather}
where the expectation is taken over the joint distribution of $P$ and $Z_{(N-k+1)}$. 

Due to the complicated nature of the distribution of the instantaneous SIR at the $k$-th best SU-Rx, it is difficult to obtain exact expressions for $E\left[R_{(N-k+1)}\right]$, $\alpha(\theta, k, N)$ and $\overline{ P_{e}}(k,N)$. Therefore, in this paper we consider another approach based on EVT to analyze the performance of the $k$-th best SU in terms of average throughput, effective throughput, outage probability and average BER.  %In what follows, we use extreme value theory to derive asymptotic analytical expressions for the average throughput, effective throughputs, average PER and outage probability of the $k$-th best SU under a continuous power adaptation shceme. 

\section{ Asymptotic Performance Analysis} 
In this section, we derive the limiting distribution of $Z_{\left(N-k+1\right)}$ in Proposition 1 below. Then we use this result to analyze the average, effective throughputs, average BER and outage probability of the $k$-th best SU. 

\subsection{ The Limiting Distribution of  $Z_{\left(N-k+1\right)}$} 

\noindent{\textbf{Proposition 1:}}  Let $Z_{(N-k+1)}$ denote the $k$-th largest order statistic of $N$ i.i.d. random variables with a common CDF of $F(z)$, as expressed in (\ref{eq:4}), then for a fixed $k$ and $N \to \infty$, $\frac{ Z_{(N-k+1)}- a}{b}$ converges in distribution to a random variable $Z$ with CDF $G^{(k)}(z)$, 
which can be characterized by an inverse gamma distribution as
\begin{eqnarray}\label{eq:11}
\begin{split}
G^{(k)}(z)=\frac{\Gamma  \left( k,{\frac {1}{z}} \right)}{(k-1)!} u(z) ,  
\end{split}
\end{eqnarray}
where $a=0$, $b=\frac{ \beta \lambda}{P_{M} \left(  \left(1-\frac{1}{N}\right)^{-\frac{1}{m}} -1 \right)} >0$ and $\Gamma(s,x)= \int_{x}^{\infty} u^{s-1} e^{-u} du$  is the upper incomplete gamma function \cite{xxx}.
Furthermore, the PDF of $Z$, $f^{(k)}(z)$, can be obtained as
\begin{eqnarray}\label{eq:12}
f^{(k)}(z)= \frac{e^{- z^{-1}}}{z^{k+1} (k-1)!} u(z).  
\end{eqnarray} 

\noindent \textit{Proof}: We first investigate the limiting distribution of $Z_{(N)}$, which denotes the first largest order statistic of $N$ i.i.d. random variables. From  Proposition 2 of  \cite{6924725}, $\frac{ Z_{(N)}- a}{b}$ converges in distribution to a unit Fr\'echet distribution i.e.,
\begin{eqnarray}\label{eq:13}
G(z)= e^{-z^{-1}} u(z),
\end{eqnarray}
 where $a=0$ and $b = { F^{-1}\left(1-\frac{1}{N}\right)}=\frac{ \beta \lambda}{P_{M} \left(  \left(1-\frac{1}{N}\right)^{-\frac{1}{m}} -1 \right)}$. 

Applying Proposition 1 of \cite{8269400} with $G(z)$  as in (\ref{eq:13}), it follows that for a fixed $k$ and $N \to \infty$, the sequence $\frac{ Z_{(N-k+1)}}{b}$ converges in distribution to a random variable $Z$ with CDF of $G^{(k)}(z)$, which can be expressed in terms of $G(z)$  as
\begin{eqnarray}\label{eq:14}
\begin{split}
G^{(k)}(z)&=G(z) \sum_{j=0}^{k-1} \frac{\left[ -\log \left( G(z) \right) \right]^{j}}{j !}\\
&= e^{-z^{-1}}\sum_{j=0}^{k-1} \frac{(z^{-1})^j }{j!} u(z).\\
%&=\frac{\Gamma  \left( k,{\frac {1}{z}} \right)}{(k-1)!} u(z), 
\end{split}
\end{eqnarray}
Using the fact that $\Gamma(k,x)= (k-1)! \ e^{-x}\sum_{j=0}^{k-1} \frac{x^j }{j!}$ for an integer $k$, $G^{(k)}(z)$ can be finally expressed as in (\ref{eq:11}). By differentiating (\ref{eq:11}) we obtain (\ref{eq:12}). 

It should be noted here that Proposition 1 of \cite{8269400} can be applied for different CDF functions. In this paper we focus on the case when $G(z)$ represents a Fr\'echet CDF. In this case $Z_{(N-k+1)}$ has a limiting distribution of inverse gamma as shown in (\ref{eq:11}). This is different from what was obtained in \cite{8269400}, where Proposition 1 of \cite{8269400} was applied for the case when $G(z)$ represents Gumbel CDF and thus $Z_{(N-k+1)}$ has a limiting distribution of $\textit{Log-Gamma}$.

\subsection{ The Distribution of the ST Transmit Power} 
We consider a continuous power adaptation scheme in which the transmit power of the ST can be adapted with a power limit of $P_{S}$; therefore, the instantaneous transmit power of the ST is $P=\min \left(P_{S}, \frac{T}{|h_{0}|^{2}}\right)$. Furthermore, we consider a continuous power adaptation scheme in which the transmit power of the ST can be adapted without any power limit, i.e., $P_{S}=\infty$ \cite{7890994},  \cite{6134707}. In such case, the ST transmit power, $P$, can be written as $P=\frac{T}{|h_{0}|^{2}}$. % We refer to these   %Using the result from Proposition 1, we analyze the average and effective throughputs of the $k$-th best secondary user with limited ST power in the following proposition. 
We focus next on the PDF of the instantaneous transmit power of the ST, $P=\min \left(P_{S}, \frac{T}{|h_{0}|^{2}}\right)$, then we use this PDF and Proposition 2 to evaluate the average and effective throughputs of the $k$-th best SU. 

Considering $P=\min \left(P_{S}, X\right)$ is a continuous random variable, where  $X= \frac{T}{|h_{0}|^{2}}$ and $P_{S}$ is constant, then the CDF of the random variable $P$, $F_{P}(t)$, can be given as 
\begin{gather}\label{eq:15}
\begin{split}
F_{P}(t)= F_{X}(t)+ u(t-P_{S})- u(t-P_{S}) F_{X}(t), 
\end{split}
\end{gather}
where $F_{X}(t)$ is the CDF of  the random variable $X= \frac{T}{|h_{0}|^{2}}$ and $ u(t-P_{S})$ is the unit step function.  Then it follows that the PDF of  $P$, $f_{P}(t)$, can be expressed as 
%given by 
%\begin{gather} \label{eq:17}
%\begin{split}
% u(t-P_{S})=
%\begin{cases}
%1,\ \    t \geq P_{S} \\
%0, \ \   t < P_{S} 
%%\frac{1}{\rho} \exp\left(\frac{n}{\rho}\right) \Gamma \left( -A,\frac{n}{\rho} \right) \left(\frac{n}{\rho}\right)^{A}, \ n=1,2,... K, \\
%\end{cases} .
%\end{split}
%\end{gather}
\begin{gather}\label{eq:16}
\begin{split}
f_{P}(t)=& f_{X}(t) \left[ 1-u(t-P_{S}) \right]+ \mathcal \delta(t-P_{S}) \left[ 1- F_{X}(P_{S}) \right], 
\end{split}
\end{gather}
where $f_{X}(t)$ is the PDF of  the random variable $X$ and $\mathcal \delta(t-P_{S})$ is the Dirac delta function, the derivative of $ u(t-P_{S})$.
 
Using the PDF of $|h_{0}|^{2}$, $f_{0}(x) = \eta e^{-\eta x} u(x), $ and variable transformation then it follows that  $F_{X}(t)= e^{-\frac{\eta T}{t}} u(t)$ and  $f_{X}(t)=\frac{\eta T}{t^2} e^{-\frac{\eta T}{t}} u(t)$. Finally we can write 
\begin{gather}\label{eq:17}
\begin{split}
f_{P}(t)=&{\frac {\eta\,T}{{t}^{2}}{{ e}^{-{\frac {\eta\,T}{t}}}}}
 \left[ 1-u(t-P_{S}) \right]+ \mathcal \delta(t-P_{S}) \left[1-{{ e}^{-{\frac {\eta\,T}{P_{S}}}}}\right]. 
\end{split}
\end{gather}

\subsection{ Average and Effective Throughputs}

\noindent{\textbf{Proposition 2:}} The average and effective throughputs of the $k$-th best SU for continuous power adaptation with limited ST power are respectively given by
\begin{gather}\label{eq:18}
\begin{split}
E\left[{R_{(N-k+1)}}\right]\approx& \frac{\ln  \left( bP_{S}\right)  -{\it E{1}} \left({\frac {
\eta\,T}{P_{S}}} \right)  - \psi \left( k \right)}{ \ln(2)}, 
\end{split}
\end{gather}

\begin{gather}\label{eq:19}
\begin{split}
\alpha (\theta, k, N)\approx -\frac{1}{A} \log_{2} \left(  \frac {\Gamma  \left( k+A \right) \Gamma  \left( A+1,{\frac {\eta\,T}{P_{S}}} \right)  }{ {\left(b\eta T\right)}^{A}\left( k-1 \right) !}  \right.\\ 
\left.+ {\frac { \Gamma  \left( k+A \right) \left( 1-{{ e}^{-{\frac {\eta\,T}{P_{S}}}}} \right) }{ \left( bP_{S} \right) ^{A} \left( k -1 \right) !} }\right), 
\end{split}
\end{gather}
for fixed $k$ and $ N \to \infty$, where $E_{1}(x)= \int_{x}^{\infty} \frac{e^{-y}}{y} dy, x >0 $ is the exponential integral function \cite [Eq. (5.1.4)]{abramowitz1964handbook}, $\Gamma(s,x)= \int_{x}^{\infty} u^{s-1} e^{-u} du$  is the upper incomplete gamma function \cite[Eq. (8.350.2)]{xxx} $\psi(x)$ is the digamma function \cite[Eq. (8.360.1)]{xxx}. %and  $E_{0}=-\psi(1)=0.5772156649$  is the Euler constant. 

\noindent{\textit{Proof:}} 
Average Throughput:

From Proposition 1, the CDF of $\frac{ Z_{(N-k+1)}}{b}$ approaches the CDF of $Z$ for a fixed $k$ and $N \to \infty$, where the CDF of $Z$ is as expressed in (\ref{eq:11}). Or equivalently, the PDF of $ Z_{(N-k+1)}$ can be approximated by the PDF of $ b Z$  for a fixed $k$ and $N \to \infty$, where the PDF of $Z$ is as expressed in (\ref{eq:12}). Then for  a fixed $k$, $N \to \infty$ and conditioning on the ST transmit power $P$, $E\left[{R_{(N-k+1)}}|P\right]$ can be approximated as 
\begin{gather}\label{eq:20}
\begin{split}
% &=%\frac{1}{\ln(2)} E\left[\log\left(1+P_{j} Z_{(N-k+1)} \right) |P_{j} \right]\\
E\left[{R_{(N-k+1)}|P}\right] &\approx\frac{1}{\ln(2)} E\left[\log\left(1+ b P  Z \right) |P\right]\\
& = \frac{1}{\ln(2)} \int_{0}^{\infty} \ln(1+bP z) \frac{ e^{- z^{-1}} }{ z^{k+1} (k-1)!}dz.  \\
%&\approx  \frac{1}{\ln(2)} \int_{0}^{\infty} \ln(bP_{j} z) \frac{ e^{- z^{-1}} }{ z^{k+1} (k-1)!}, 
\end{split}
\end{gather}
Noting that $b$ is an increasing function of $N$, we have $\ln(1+bP z) \approx \ln(bP z)$ in (\ref{eq:20}) for large $N$. Using this and variable transformation of $u=(bP z)^{-1}$, $E\left[{R_{(N-k+1)}}|P\right]$ can be further approximated as
\begin{gather}\label{eq:21}
\begin{split}
E\left[{R_{(N-k+1)}}|P\right] &\approx  \int_{0}^{\infty}  \frac{-\ln(u)\left(bP\right)^{k}  e^{-bPu} u^{k-1} }{ \ln(2)  (k-1)!}du \\
&=\frac{ \ln(bP) -\psi(k)}{\ln(2)}, \\
\end{split}
\end{gather}
where the above integral is evaluated with help of  \cite[Eq. (4.352. 1) ]{xxx}.  Averaging $\ln(P)$ over the PDF of $f_{P}(t)$ in (\ref{eq:17}) yields
\begin{gather}\label{eq:22}
\begin{split}
 \int_{0}^{\infty}\ln(t) f_{P}(t) dt=& \int_{0}^{P_{S}}\ln(t)  \frac{\eta T}{t^2} e^{-\frac{\eta T}{t}}dt \\
&+\ln \left( P_{S}\right)  \left[1-{ e}^{-{\frac {\eta\,T}{ P_{S}}}}\right]. 
\end{split}
\end{gather}
Using variable transformation  of $u=P_{S} t^{-1}$ with help of  \cite[ Eq. (4.331. 2) ]{xxx} and after some basic algebraic manipulation, we have 
\begin{gather}\label{eq:23}
\begin{split}
 \int_{0}^{P_{S}}\ln(t)  \frac{\eta T}{t^2} e^{-\frac{\eta T}{t}}dt ={{ e}^{-{\frac {\eta\,T}{P_{S}}}}}\ln  \left( P_{S} \right) -{\it E_{1}}  \left({\frac {\eta\,T}{P_{S}}} \right).   
\end{split}
\end{gather}
Combining (\ref{eq:21}), (\ref{eq:22})  and (\ref{eq:23}),  it follows that $E\left[{R_{(N-k+1)}}\right]$ is as expressed in (\ref{eq:18}).

Effective Throughput: 
Conditioning on the ST transmit power $P$ in (\ref{eq:8}) and by exploiting  Lemma 2 of the Appendix,  we infer that $E\left[ \left( 1+ P  Z_{(N-k+1)} \right)^{-A} | P \right]$  can be approximated as 
\begin{gather} \label{eq:24}
\begin{split}
 E\left[ \left( 1+ P  Z_{(N-k+1)} \right)^{-A} | P \right] &\approx  E\left[ \left( 1+ bP  Z \right)^{-A} | P \right]\\
& = \int_{0}^{\infty}   \frac{\left(  1+bP z \right)^{-A}  e^{-z^{-1}} }{ z^{k+1} (k-1)!} dz,\\
\end{split}
\end{gather}
for fixed $k$ and $N\to \infty$. Making use as above of $ 1+bP z \approx bP z$ for large $N$ in (\ref{eq:24})  and variable transformation of $u=(bP z)^{-1}$, $E\left[ \left( 1+ P  Z_{(N-k+1)} \right)^{-A} | P \right]$ can be further approximated as
\begin{gather} \label{eq:25}
\begin{split}
 E\left[ \left( 1+ P  Z_{(N-k+1)} \right)^{-A} | P \right]& \approx \int_{0}^{\infty}   \frac{ \left(bP\right)^{k} u^{A+k-1} e^{-bP u} }{(k-1)!} dz\\ %\approx \int_{0}^{\infty}   \frac{\left(bP z \right)^{-A}  e^{-z^{-1}} }{ z^{k+1} (k-1)!} dz\\
&={\frac {{\left( bP\right)}^{-A}\Gamma  \left( A+k \right) }{(k-1)!}},
\end{split}
\end{gather}
where the above integral is evaluated with help of \cite [Eq. (3.381.4)] {xxx}. Averaging (\ref{eq:25}) over the PDF of $f_{P}(t)$ in (\ref{eq:17}) yields
\begin{gather}\label{eq:26}
\begin{split}
 E\left[ \left( 1+ P  Z_{(N-k+1)} \right)^{-A} \right] \approx&  \int_{0}^{\infty} {\frac {\left(bt\right)^{-A}\Gamma  \left( A+k \right) }{(k-1)!}} f_{P}(t) dt\\
=& \int_{0}^{P_{S}}  {\frac { \Gamma  \left( k+A \right) \eta\,T { e}^{-{\frac {\eta\,T}{t}}}}{
b^{A} {t}^{A+2} \left( k-1 \right) !\,}}dt \\
&+{\frac { \Gamma  \left( k+A \right) \left( 1-{{ e}^{-{\frac {\eta\,T}{P_{S}}}}} \right) }{ \left( bP_{S} \right) ^{A} \left( k
-1 \right) !} }.
\end{split}
\end{gather}
Using variable transformation  of $u=\eta T t^{-1}$ and using the definition of the upper incomplete gamma function, $\Gamma(s,x)= \int_{x}^{\infty} u^{s-1} e^{-u} du$, we have 
\begin{gather}\label{eq:27}
\begin{split}
 \int_{0}^{P_{S}}  {\frac { \Gamma  \left( k+A \right) \eta\,T { e}^{-{\frac {\eta\,T}{t}}}}{
b^{A} {t}^{A+2} \left( k-1 \right) !\,}} dt = \frac {\Gamma  \left( k+A \right) \Gamma  \left( A+1,{\frac {\eta\,T}{P_{S}}} \right)  }{ {\left(b\eta T\right)}^{A}\left( k-1 \right) !}. 
\end{split}
\end{gather}
Combining (\ref{eq:26}), (\ref{eq:27}) and (\ref{eq:8}), it follows that $\alpha (\theta, k, N)$ is as expressed in (\ref{eq:19}). 

While we focused in Proposition 2 on analyzing the average and effective throughputs under the limited ST power adaptation, i.e., $P=\min \left(P_{S}, \frac{T}{|h_{0}|^{2}}\right)$, it should be noted that simpler expressions can be obtained under the unlimited ST power case. i.e., $P=\frac{T}{|h_{0}|^{2}}$. These expressions are useful when $P_{S}\to \infty$ \cite{6134707}, \cite{7890994} and they serve as upper bounds on the average and effective throughput under the limited ST power case. % Now we consider a continuous power adaptation with $P_{S}=\infty$ \cite{6134707}, \cite{7890994}. In this case, the ST transmit power, $P$, can be written as $P=\frac{T}{|h_{0}|^{2}}$.
Using the result from Proposition 2, we derive the average and effective throughputs of the $k$-th best SU with unlimited ST power in the following corollary. \\ 
 
\noindent{\textbf{Corollary 1:}} The average and effective throughputs of the $k$-th best SU for continuous power adaptation with unlimited ST power are respectively given by
\begin{gather}\label{eq:28}
\begin{split}
E\left[{R_{(N-k+1)}}\right]\approx&\frac{\ln(bT\eta)-\psi(k)+E_{0} }{\ln(2)}, 
\end{split}
\end{gather}

\begin{gather}\label{eq:29}
\begin{split}
\alpha (\theta, k, N)\approx\frac{\ln(bT\eta)}{\ln(2)} -\frac{1}{A} \log_{2} \left( {\frac {\Gamma \left( A+k \right) \Gamma \left( A+1 \right) }{ \left( k -1 \right) ! }}\right),  
\end{split}
\end{gather}
for fixed $k$ and $ N \to \infty$, where $E_{0}=-\psi(1)=0.5772156649$  is the Euler constant.  

\noindent{\textit{Proof:}}

Average Throughput: Using Puiseux series for the exponential integral function, $E_{1}(x)$,
 we have
\begin{gather}\label{eq:30}
\begin{split}
E_{1}(x)= -E_{0}- \ln(x)-  \sum_{n=1}^{\infty} \frac{(-x)^n}{n n!}, \ x>0.
\end{split}
\end{gather}
Invoking  (\ref{eq:18}) and with the help of (\ref{eq:30}),  one can show  that  $\lim_{P_{S} \to \infty}  \left( \ln\left( bP_{S}\right)  -{\it E{1}} \left({\frac {\eta\,T}{P_{S}}} \right) \right)=\ln\left(b  T\eta \right)+E_{0}$. Therefore, as $P_{S} \to \infty$, the average throughput is as expressed in (\ref{eq:28}).

Effective Throughput: Invoking (\ref{eq:19}) and $\lim_{P_{S} \to \infty} \Gamma(A+1,\frac{\eta T}{P_{S}})= \Gamma(A+1)$ one can show that the effective throughput is as expressed in (\ref{eq:29})  as $P_{S} \to \infty$. 

\subsection{ Average BER} 
We now derive the average BER for the limited and unlimited continuous ST power in the following proposition. 

\noindent{\textbf{Proposition 3:}}
 The average BER of the $k$-th best SU for continuous limited power adaptation scheme can be approximated as 
\begin{gather} \label{eq:31}
\begin{split}
\overline{ P_{e}}(k,N) \approx & %\int_{0}^{\infty}  {\frac { 2c\left( gb t \right) ^{k/2}{{\sl K}_{k}\left(2\,\sqrt {gbt} \right)}}{ \left( k-1 \right) !}} f_{P}(t) dt\\
\int_{0}^{P_{S}}\!2\,{\frac { \left( v b t \right) ^{k/2}{{\sl K}_{k}\left(2
\,\sqrt {v b t}\right)}\eta\,T}{ \left( k-1 \right) !\,{t}^{2}}{{ e}^
{-{\frac {\eta\,T}{t}}}}}\,{\rm d}t\\
&+{\frac {2 \left( v b P_{S} \right) ^{k
/2}{{\sl K}_{k}\left(2\,\sqrt {v b P_{S}}\right)}}{ \left( k-1 \right) !}
 \left( 1-{{ e}^{-{\frac {\eta\,T}{P_{S}}}}} \right) },
\end{split}
\end{gather}
 for fixed $k$ and $ N \to \infty$, where ${\sl K}_{\nu}\left(\cdot \right)$ is the modified Bessel function of the second kind and order $\nu$ \cite [Eq. (8.407.1)]{abramowitz1964handbook}
Furthermore, for the unlimited  ST transmit power, the average BER of the $k$-th best SU can be approximated as
\begin{gather} \label{eq:32}
\begin{split}
\overline{ P_{e}}(k,N) \approx {\frac { c\left( \eta\,T  vb \right) ^{k/2-1} G^{3, 0}_{0, 3}\left(\eta\,Tvb\, \Big\vert\,^{-}_{1+k/2, 2-k/2, 1-k/2}\right)
}{  \left( k-1 \right)! }}, 
\end{split}
\end{gather}
for fixed $k$ and $ N \to \infty$, where $G_{p,q} ^{m,n}(.)$ is the Meijer G-function \cite{prudnikov1990integrals}. 

\noindent{\textit{Proof:}} 
Conditioning on the ST transmit power $P$ in (\ref{eq:10}) and by exploiting  Lemma 1 of the Appendix,  we infer  that  $E\left[ e^{-v PZ_{(N-k+1)}}\right| P]$ can be approximated as 
\begin{gather} \label{eq:33}
\begin{split}
 E\left[ e^{-v P Z_{(N-k+1)}}\right| P] &\approx E\left[ e^{-v b P Z}\right| P]\\
& = \int_{0}^{\infty} \frac {{{e}^{-v b P z}}{{ e}^{- {z}^{-1}}}}{{z}^{k+1} \left( k-1 \right) !} dz\\
&={\frac { 2\left( v b P \right) ^{k/2}{{\sl K}_{k}\left(2\,\sqrt {v b P} \right)}}{ \left( k-1 \right) !}}, 
%&\approx  p_{0}+E\left[ \left( 1+ bP_{j} Z_{(N-k+1)} \right)^{-A} |P_{j} \right] \\
%&=\int_{0}^{\infty}   \frac{\left( 1+ C_{j} Z \right)^{-A}  e^{-z^{-1}} }{ z^{k+1} (k-1)!} dz .  
\end{split}
\end{gather}
for a fixed $k$ and $N \to \infty$, where the above integral is evaluated with help of \cite [Eq. (2.11)]  {Glasser12theintegrals}. It is hard to find an analytical expression for the average BER for the limited ST transmit power case. Therefore, averaging  (\ref{eq:33}) over  $f_{P}(t)$ in  (\ref{eq:17}) yields the average BER for limited ST power as in (\ref{eq:31}). For the unlimited  ST transmit power, one can show that by letting $P_{S} \to \infty$ in (\ref{eq:31}), we have %PDF of $P=\frac{T}{|h_{0}|^{2}}$, we have %can be expressed as $f_{P}(t)={\frac {\eta\,T}{{t}^{2}}{{ e}^{-{\frac {\eta\,T}{t}}}}} \mathcal U(t)$. Therefore, averaging  (\ref{eq:34}) over  $f_{P}(t)={\frac {\eta\,T}{{t}^{2}}{{ e}^{-{\frac {\eta\,T}{t}}}}} \mathcal U(t)$, yields 
\begin{gather} \label{eq:34}
\begin{split}
\overline{ P_{e}}(k,N) \approx \int_{0}^{\infty}  {\frac { 2c\left( v b t \right) ^{k/2}{{\sl K}_{k}\left(2\,\sqrt {v bt} \right)}}{ \left( k-1 \right) !}} {\frac {\eta\,T}{{t}^{2}}{{ e}^{-{\frac {\eta\,T}{t}}}}}dt. 
\end{split}
\end{gather}
The above integral can be expressed in terms of the Meijer G-function as in (\ref{eq:32}). It should be noted that the Meijer G-function can be easily and efficiently computed using the most standard software packages like MAPLE and MATHEMATICA. 

\subsection{Outage Probability}
We now derive the outage probability for limited and unlimited continuous ST power in the following proposition. 

\noindent{\textbf{Proposition 4:}}
 The outage probability of the $k$-th best SU for continuous limited power adaptation scheme can be approximated as 
\begin{gather} \label{eq:35}
\begin{split}
P_{out}(x_{0}) \approx & \int_{0}^{P_{S}}\!{\Gamma 
 \left( k,{\frac {bt}{x_{0}}} \right) \frac {\eta\,T}{ \left( k-1 \right) !\,{t}^{2}} {{ e}^{-{\frac {\eta\,T}{t}}}}}
\,{d}t\\
&+{\Gamma  \left( k,{\frac {b
P_{S}}{x_{0}}} \right)  \frac{\left( 1-{{ e}^{-{\frac {\eta\,T}{P_{S}}}}} \right)}{\left( k-1 \right) !} },
\end{split}
\end{gather}
for fixed $k$ and $ N \to \infty$. Furthermore, for the unlimited  ST transmit power, the outage probability of the $k$-th best SU can be approximated as
\begin{gather} \label{eq:36}
\begin{split}
P_{out}(x_{0})  \approx \frac{ 2 \left( \frac{\eta T b}{x_{0}}\right)^\frac{k}{2} {\sl K}_{k}\left(2\,\sqrt { \frac{\eta T b}{x_{0}}} \right)}{{\left( k-1 \right) !} }, 
\end{split}
\end{gather}
for fixed $k$ and $ N \to \infty$.

\noindent{\textit{Proof:}}  The outage probability of the $k$-th best SU, $P_{out}(x_{0})$, can be expressed as
\begin{gather} \label{eq:37}
\begin{split}
P_{out}(x_{0})&= \int_{0}^{\infty}  {\rm Pr} \{ t Z_{\left(N-k+1\right)} \leq x_{0}\} f_{P}(t) dt\\
&= \int_{0}^{\infty}  {\rm Pr} \{ t Z_{\left(N-k+1\right)} \leq x_{0}\} f_{P}(t) dt.\\
\end{split}
\end{gather}
 where $f_{P}(t)$ as given in (\ref{eq:17}). From Proposition 1, the CDF of $\frac{ Z_{(N-k+1)}}{b}$ approaches  the CDF of $Z$ for fixed $k$ and $N \to \infty$, where the CDF of $Z$ is as expressed in (\ref{eq:11}). Then, we have
\begin{gather} \label{eq:38}
\begin{split}
P_{out}(x_{0})&= \int_{0}^{\infty}  {\rm Pr} \{ t Z_{\left(N-k+1\right)} \leq x_{0}\} f_{P}(t) dt\\
&= \int_{0}^{\infty}  {\rm Pr} \left \{ \frac{ Z_{\left(N-k+1\right)}}{b} \leq \frac{x_{0}}{bt} \right \} f_{P}(t) dt\\
&\approx \int_{0}^{\infty}  {\rm Pr} \left \{ Z \leq \frac{x_{0}}{bt} \right \} f_{P}(t) dt\\
&= \int_{0}^{\infty}  \frac{\Gamma  \left( k,{\frac {bt}{x_{0}}} \right)}{(k-1)!} f_{P}(t) dt.
\end{split}
\end{gather}
 Making use of (\ref{eq:17}) in (\ref{eq:38}), the outage probability for the limited ST power is as in (\ref{eq:35}). For the unlimited  ST transmit power, one can show that by letting $P_{S} \to \infty$ in (\ref{eq:35}), we have

\begin{gather} \label{eq:39}
\begin{split}
P_{out}(x_{0}) \approx & \int_{0}^{\infty}\!{\Gamma 
 \left( k,{\frac {bt}{x_{0}}} \right) \frac {\eta\,T}{ \left( k-1 \right) !\,{t}^{2}} {{ e}^{-{\frac {\eta\,T}{t}}}}}
\,{d}t\\
&= \frac{ 2 \left( \frac{\eta T b}{x_{0}}\right)^\frac{k}{2} {\sl K}_{k}\left(2\,\sqrt { \frac{\eta T b}{x_{0}}} \right)}{{(k-1)!}}, 
\end{split}
\end{gather}
where the integral above is evaluated with help of \cite[Eq. (6.453)] {xxx} after variable transformation of $u=(\eta Tt)^{-1}$. 
%\begin{comment}

\subsection{ Effect of Imperfect CSI } 
In practical environments, the ST has only a partial channel knowledge of the ST to PR channel, $h_{0}$. In this case, the CSI on $h_{0}$ provided to the ST is outdated due to the time-varying nature of the wireless link \cite{6094132}. The outdated CSI can be described using the correlation model as \cite{6094132}.
\begin{gather} \label{eq:A}
\begin{split}
h_{0}= \rho \hat{h}_{0} +\sqrt { 1-\rho^{2}} \tilde{h}_{0}, 
\end{split}	
\end{gather}
where $\hat{h}_{0}$ is the outdated channel information available at the ST, $\tilde{h}_{0}$ is a complex Gaussian random variable with zero mean and unit variance, and uncorrelated with $h_{0}$. The correlation coefficient $\rho$ ($0 \leq \rho \leq 1$) is a constant, which is used to evaluate the impact of channel estimation error and feedback delay on the CSI \cite{6094132}. It is assumed that the ST knows the outdated channel information $\hat{h}_{0}$ and the correlation coefficient $\rho$ as well. In view of $|h_{0}|^{2}$ being an exponentially distributed random variable with parameter $\eta$, the estimated channel power gain $|\hat{h}_{0}|^{2}$ is also an exponentially distributed RV with parameter $\hat{\eta}$, where  
$\eta^{-1}=\rho^{2} \hat{\eta}^{-1} +\left( 1-\rho^{2} \right) $.

As we discussed in Section II, when the ST has a perfect CSI of $h_{0}$, it can access the spectrum if the peak interference power constraint can be satisfied. However, it is hard to satisfy the instantaneous interference constraint at the PR if only the outdated CSI is available at ST \cite{6094132}. Therefore, a more flexible constraint based on a pre-selected interference outage probability is adopted \cite{6094132}, %\cite{6880852},
\cite{6307791}. Considering the imperfect CSI effect, the transmit power of the ST in (\ref{eq:FFFF}) can be rewritten as \cite{6094132}
\begin{equation} \label{eq:B}
P=\min \left(P_{S}, r_{I}\frac{T}{|\hat{h}_{0}|^{2}}\right),
\end{equation}
where $ r_{I}$ denotes the power margin factor which can be expressed as \cite{6094132}
\begin{equation} \label{eq:C}
\begin{split}
r_{I}=&(-1+2 \rho^{2}) \\ 
&+ \frac{ 1-\rho^{2}- (1-2 \Gamma_{0}) \sqrt{(1-\rho^{2}) \left(1-(1-2 \Gamma_{0})^2 \rho^{2}\right) }}{2 \Gamma_{0} (1-\Gamma_{0})}, 
\end{split}
\end{equation}
where $\Gamma_{0}$ denotes the predetermined interference outage probability. As a special case, a power margin factor of $r_{I}=1 $, i.e., ($\rho=1$) indicates the perfect CSI of $h_{0}$ and therefore the ST transmit power in (\ref{eq:B}) reduces to (\ref{eq:FFFF}).

 For further practical considerations, we address the imperfect CSI of the channels in the secondary network, $h_{i}$, for $i=1, 2, ..., N$. The outdated CSI can be described as 
\begin{gather} \label{eq:5000}
\begin{split}
h_{i}= \delta \hat{h}_{i} +\sqrt { 1-\delta^{2}} \  \tilde{h}_{i}, \ \ \  i=1, 2, ..., N
\end{split}	
\end{gather}
where $\hat{h}_{i}$ is the outdated channel information of the $i$-th secondary link available at the ST, $\tilde{h}_{i}$ is a complex Gaussian random variable with zero mean and unit variance, and uncorrelated with $h_{i}$. The correlation coefficient $\delta$ ($0 \leq \delta  \leq 1$) is a constant that  describes  the impact of outdated CSI.   In view of $|h_{i}|^{2}$ being a Gamma distributed random variable with parameters $m$ and $\beta$, the estimated channel power gain $|\hat{ h}_{i}|^{2}$ is also a Gamma distributed random variable with parameters $m$ and $\hat{\beta}$. 

 It should be noted that the expressions derived for average throughput, effective throughput, average BER and outage probability in previous subsections assuming a perfect CSI hold for the imperfect CSI case after replacing $\eta$ with $\hat{\eta}$ and $T$ with $\ r_{I} \, T$, due to the imperfect CSI on $h_{0}$. And replacing $b$ with $\hat{b}$ due to the imperfect CSI on $h_{i}$, where $\hat{b}=\frac{ \hat{\beta} \lambda}{P_{M} \left(  \left(1-\frac{1}{N}\right)^{-\frac{1}{m}} -1 \right)} >0$.

\section{ Numerial results } 
\subsection{ Perfect CSI}
In this subsection, we numerically illustrate and verify the obtained asymptotic expressions in Section III under the perfect CSI condition,  which refers to $\rho=1$ with ST transmit power is as in (\ref{eq:FFFF}) as described at the end of Section III. \textit {F} . 

In Fig. 2, we plot the average throughput of the $k$-th best SU versus the number of secondary users, $N$, for unlimited ST power, $P_{S}=\infty$, and limited ST power with $P_{S}=10\ \text{dB}$, for $k=1,2,3$. We validate the obtained asymptotic expressions for the average throughput using Monte Carlo simulations. We observe that the accuracy increases as $N$ increases. Furthermore, we observe that the asymptotic results are accurate for not so large (realistic) values of $N$. For example, $N=40$ is considered sufficiently large to confirm the accuracy of the asymptotic results compared to simulations. This suggests that the EVT is a powerful approach that approximates the performance for realistic and large values of $N$ as well.  In Fig. 3, we plot the average throughput of the best SU versus the ST power, $P_{S}$, in $\text{dB}$. We observe that, compared to the simulations, the accuracy of the asymptotic average throughput increases as $N$ increases from 6 to 30. We also observe that the accuracy of the asymptotic average throughput increases as $P_{S}$ increases. Furthermore, for larger values of $P_{S}$ the asymptotic average throughput approaches the one with unlimited ST power, $P_{S}=\infty$. 

\begin{figure}\label{fig:2}
\begin{center}
\includegraphics[width=1\columnwidth]{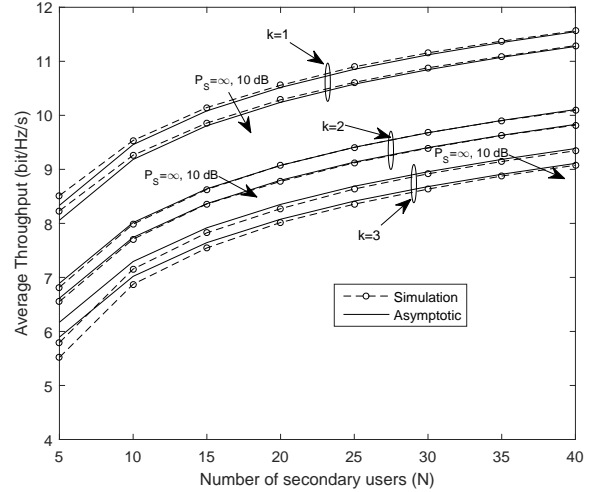}
\caption{Average throughput of the $k$-th best SU versus the number of secondary users $N$ with unlimited ST power, $P_{S}=\infty$ and limited ST power with $P_{S}=10\  \text{dB}$, for $k=1,2,3$,   $\lambda=2$, $\beta=3$, $\eta=20$, $m=2$, $T=-10$ \text{dB} and $P_{M}=0 $ \text{dB}.  }
\end{center}
\end{figure} 

\begin{figure}\label{fig:3}
\begin{center}
\includegraphics[width=1\columnwidth]{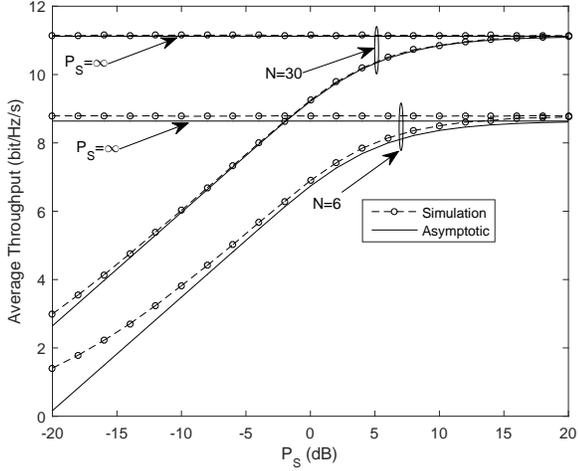}
\caption{Average throughput of the best SU versus the ST power, $P_{S} (\text{dB})$, for $N=6, 30$, $\lambda=2$, $\beta=3$, $\eta=20$, $m=2$, $T=-10$ \text{dB}, $P_{M}=0 $ \text{dB}. }
\end{center}
\end{figure}

\begin{figure} \label{fig:4}
\begin{center}
\includegraphics[width=1\columnwidth]{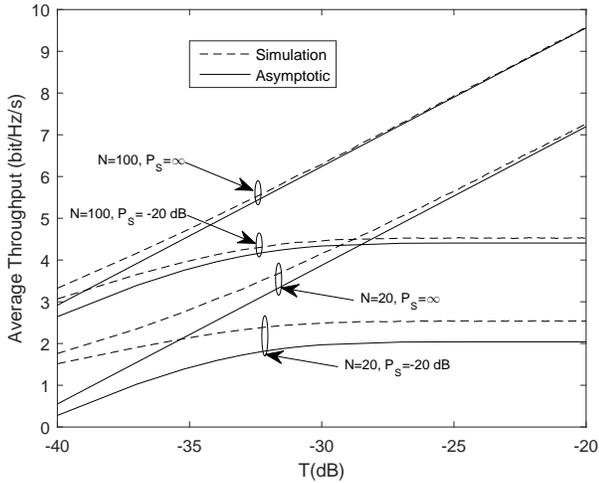}
\caption{Average throughput of the best SU versus interference level, $T$, in $ (\text{dB})$, for $N=20, 100$, $\lambda=2$, $\beta=3$, $\eta=20$, $m=2$ and $P_{M}=0 $ \text{dB} at $P_{S}= -20 \ \text{dB}$, $P_{S}= \infty$.   }
\end{center}
\end{figure}

In Fig. 4, we plot the average throughput of the best SU versus the interference level, $T$, in $\text{dB}$ for unlimited ST power, $P_{S}=\infty$, and limited ST power with $P_{S}=-20\ \text{dB}$. Some interesting observations can be made from this figure. First, we observe that, compared to the simulations, the accuracy of the asymptotic average throughputs increase as $N$ increases from 20 to 200. Second, as $T$ or $P_{S}$ increases, the accuracy of the asymptotic average throughputs also increases. Last, for the limited ST power, $P_{S}=-20 \text{dB} $, the average throughput is saturated and it does not improve as $T \geq -30 \ \text{dB}$. This is due to the fact that for higher values of $T$, the ST will select $P_{S}$ with a higher probability.  

\begin{figure} \label{fig:5}
\begin{center}
\includegraphics[width=1\columnwidth]{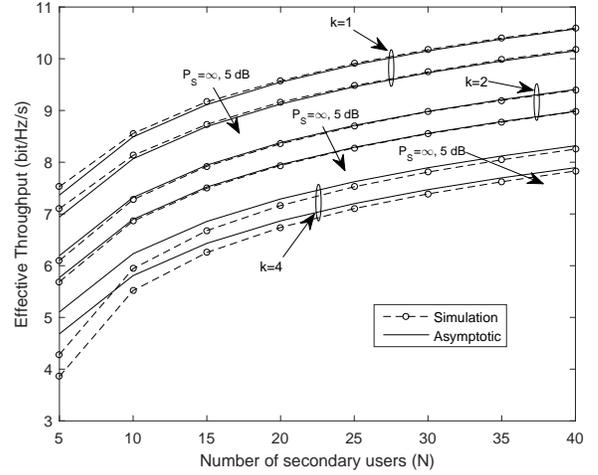}
\caption{Effective throughput of the $k$-th best SU versus the number of secondary users $N$ with unlimited ST power, $P_{S}=\infty$ and limited ST power with $P_{S}=5\  \text{dB}$, for $k=1,2,4$, $A=1/2$, $\lambda=2$, $\beta=3$, $\eta=20$, $m=2$, $T=-10$ \text{dB} and $P_{M}=0 $ \text{dB}.}
\end{center}
\end{figure}

In Fig. 5, we plot the effective throughput of the $k$-th best SU versus the number of secondary users, $N$, with unlimited ST power, $P_{S}=\infty$ and limited ST power with $P_{S}=5\ \text{dB}$, for $k=1,2,4$. We observe that the accuracy of the asymptotic effective throughput increases as $N$ increases. However, it is shown that for $k=4$, the asymptotic effective throughputs is less accurate for small to moderate values of $N$. This is because the asymptotic analysis is more accurate for large $N$ relative to a fixed $k$. Consequently, if the value of $k$ is close enough to $N$, it is expected that the asymptotic expression will be less accurate. 

\begin{figure}\label{fig:6}
\begin{center}
\includegraphics[width=1\columnwidth]{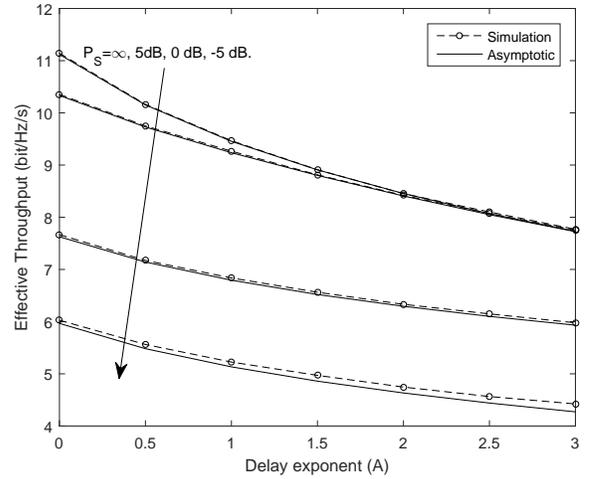}
\caption{Effective throughput of the best SU versus delay exponent at $N=30$, for $\lambda=2$, $\beta=3$, $\eta=20$, $m=2$, $T=-10\  \text{dB}$, $P_{M}=0 $ and different values of $P_{S}$.  }
\end{center}
\end{figure}

In Fig. 6, we plot the effective throughput of the best SU versus the delay exponent, $A$, for $N=30$ and different values of $P_{S}$. We observe that the effective throughput significantly decreases for smaller values of $P_{S}$. On the other hand, for reasonably large values of $P_{S}$, the effective throughput does not significantly improve compared to the unlimited ST power case, $P_{S}=\infty$. This is due to the fact that for higher values of $P_{S}$ the effective throughput is dominated by the interference level $T$. 

%\begin{figure} \label{fig:7}
%\begin{center}
%\includegraphics[width=1\columnwidth]{Fig7.eps}
%\caption{Outage probability of the $k$-th best SU versus the number of secondary users ($N$) with unlimited ST power, $P_{S}=\infty$ and limited ST power with $P_{S}=-10\  \text{dB}$,  for $k=1,2$, $\lambda=2$, $\beta=3$, $\eta=20$, $T=-20$ \text{dB}, $P_{M}=0 $ \text{dB} and $x_{0}=10$ \text{dB}.}
%\end{center}
%\end{figure}

\begin{figure}[h]\label{fig:7}
\begin{center}
\includegraphics[width=1\columnwidth]{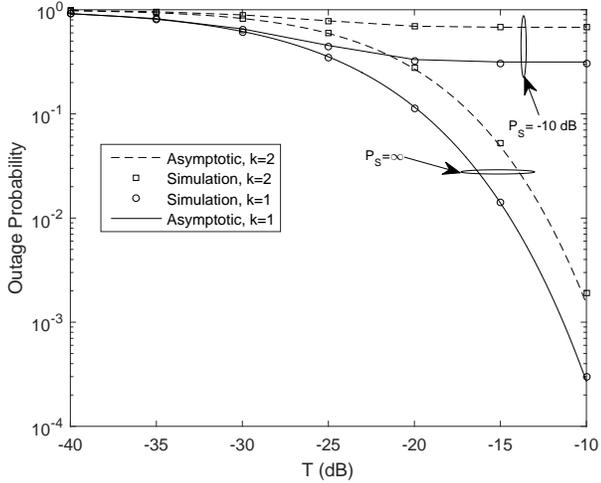}
\caption{Outage probability of the  $k$-th best SU versus interference level, $T$, in $ (\text{dB})$  at $N=30$ with unlimited ST power, $P_{S}=\infty$ and limited ST power with $P_{S}=-10\  \text{dB}$,  for $k=1,2$, $\lambda=2$, $\beta=3$, $\eta=20$, $m=2$, $P_{M}=0 $ and $x_{0}=13$ \text{dB}. }
\end{center}
\end{figure}
\begin{figure}[h]\label{fig:8}
\begin{center}
\includegraphics[width=1\columnwidth]{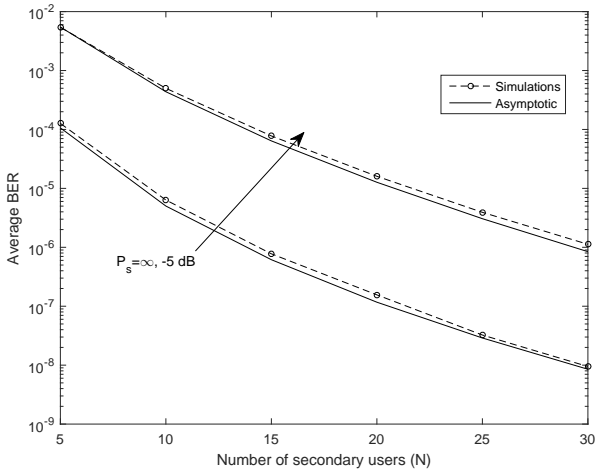}
\caption{Average BER  of BFSK ($C=g=0.5$)  versus the number of secondary users $N$ with unlimited ST power, $P_{S}=\infty$ and limited ST power with $P_{S}=-5\  \text{dB}$, for $k=1$, $\lambda=2$, $\beta=3$, $\eta=20$, $m=2$, $T=-10$ \text{dB} and $P_{M}=0 $ \text{dB}.  }
\end{center}
\end{figure}

%In Fig. 7, we plot the outage probability of the $k$-th best SU versus the number of secondary users, $N$, for the unlimited ST power, $P_{S}=\infty$ and limited ST power with $P_{S}=-10\ \text{dB}$, for $k=1,2$ and $x_{0}=10$ \text{dB} .
 In Fig. 7, the outage probability of the $k$-th best SU is plotted versus interference level, $T$, in $ (\text{dB})$ at $N=30$ with unlimited ST power, $P_{S}=\infty$ and limited ST power with $P_{S}=-10\ \text{dB}$, for $k=1,2$ at $x_{0}=13$ \text{dB}. The saturation in the outage probability for the limited ST power case is due to the fact that for higher values of $T$, the ST will select $P_{S}$ for most of the time. In Fig. 8, we plot the asymptotic average BER as a function of the number of secondary users, $N$, for the unlimited ST power with $P_{S}=\infty$ and limited ST power with $P_{S}=-5\ \text{dB}$, for $k=1$. We validate the obtained analytical results using Monte Carlo simulations. We observe that the asymptotic expression is accurate even for moderate values of $N$. We also observe that the average BER for the unlimited ST power serves as a lower bound on the average BER with limited ST power case.

%\clearpage
\subsection{Imperfect CSI} 
In this subsection, we consider the impact of imperfect CSI as described in Section III. \textit{F}.  In Fig. 9 we plot the effective throughput for $A=1/2$ and $A=0$ ( the average throughput) versus the correlation coefficient, $\rho$, at $N=40$ and  $\Gamma_{0}=10\%$ with unlimited ST power, $P_{S}=\infty$.  We observe that the effective throughput is improved as the quality of the channel estimate increases, i.e., $\rho$ increases. As expected, for $\rho=1$, which implies perfect CSI, the highest effective throughput is achieved. 
 \begin{figure}[h]\label{fig:9}
\begin{center}
\includegraphics[width=1\columnwidth]{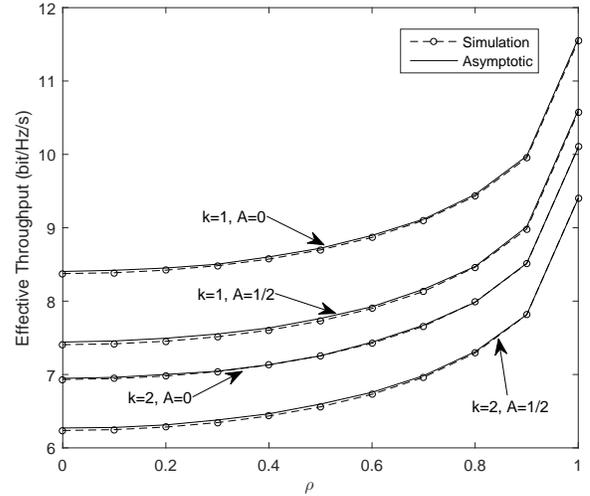}
\caption{Effective throughput of the $k$-th best SU versus the correlation coefficient, $\rho$, with unlimited ST power, $P_{S}=\infty$, $\Gamma_{0}=10\%$ for $N=40$, $k=1,2$, $A=0, 1/2$, $\lambda=2$, $\hat{\beta}=3$, $\hat{\eta}=20$, $m=2$, $T=-10$ \text{dB} and $P_{M}=0 $ \text{dB}.}
\end{center}
\end{figure}

\section{Conclusion}
We analyzed the asymptotic performance of the $k$-th best SU for an interference-limited secondary multiuser network of underlay CR systems. We used extreme value theory to show that the $k$-th highest SIR converges in distribution to an inverse gamma random variable for a fixed $k$ and large number of secondary users. We used this result to analyze the asymptotic average throughput, effective throughput, average BER and outage probability for the $k$-th best SU under continuous power adaptation at the ST. We verified the accuracy of the derived asymptotic expressions, for different system parameters, through Monte Carlo simulations.

%We analyzed the asymptotic performance of the $k$-th best SU for an interference-limited secondary multiuser network of underlay cognitive radio (CR) systems. We used extreme value theory to analyze the asymptotic average throughput, effective throughput, average bit error rate (BER) and outage probability for the $k$-th best SU under continuous power adaptation at the ST.  We verified the accuracy of the derived asymptotic expressions, for different system parameters, through Monte Carlo simulations.

\section*{Appendix }
Lemma 1 below establishes the connection between convergence in distribution and convergence in moment generating function (MGF). For a positive random variable $X$, the MGF of  is $E \left[e^{t X}\right]=  \int_{0}^{\infty} e^{tx} f(x) dx$, where  $f(x)$ is the PDF of $X$. \\

\noindent \textit{Lemma 1}: If $\frac{Z_{\left(N-k+1\right)}} {b}$ converges in distribution to a random variable $Z$ whose CDF is as in (\ref{eq:11}), then for a fixed $k$ we have
\begin{eqnarray}\label{eq:90}
\lim_{N\to\infty} E\left[ e^{t \left( \frac{ Z_{\left(N-k+1\right)}} {b} \right)} \right]=  E \left[e^{t Z}\right],  \ t < 0, 
%E\left[ e^{-t \left( \frac{ Z_{\left(N-k+1\right)}} {b} \right)} \right] \to E \left[e^{-t Z}\right],
\end{eqnarray}
 where  $E \left[e^{t Z}\right]={\frac { 2\left( -t \right) ^{k/2}{{\sl K}_{k}\left(2\,\sqrt {-t} \right)}}{ \left( k-1 \right) !}},  \ t < 0. $  %Or equivalently, $E\left[ e^{-t  Z_{\left(N-k+1 \right)}} \right] \approx \left[e^{-t b Z}\right]$, for a fixed   and $N \to \infty$. 

\noindent \textit{Proof}:  Theorem 2 of \cite{chareka2008converse} implies that if $E\left[ e^{t \left( \frac{ Z_{\left(N-k+1\right)}} {b} \right)} \right]$  exists for all $t < 0$ and  $\frac{Z_{\left(N-k+1\right)}} {b_{N}}$ converges uniformly in distribution to a random variable $Z$  with CDF as in (\ref{eq:11}) where $E \left[e^{t Z}\right]$ exists for all $t < 0$, then for a fixed $k$, $\lim_{N\to\infty} E\left[ e^{t \left( \frac{ Z_{\left(N-k+1\right)}} {b} \right)} \right]=  E \left[e^{t Z}\right]$, for all  $t < 0$.   We note that $E \left[e^{t Z}\right]$ for all  $t < 0$ exists and can be evaluated as 
\begin{gather} \label{eq:70}
\begin{split}
 E\left[ e^{t Z}\right] &=\int_{0}^{\infty} \frac {{{e}^{tz}}{{ e}^{- {z}^{-1}}}}{{z}^{k+1} \left( k-1 \right) !} dz\\
&={\frac { 2\left( -t \right) ^{k/2}{{\sl K}_{k}\left(2\,\sqrt {-t} \right)}}{ \left( k-1 \right) !}},  \ t <0, 
%&\approx  p_{0}+E\left[ \left( 1+ bP_{j} Z_{(N-k+1)} \right)^{-A} |P_{j} \right] \\
%&=\int_{0}^{\infty}   \frac{\left( 1+ C_{j} Z \right)^{-A}  e^{-z^{-1}} }{ z^{k+1} (k-1)!} dz .  
\end{split}
\end{gather}
where the above integral is evaluated with help of \cite [Eq. (2.11)]{Glasser12theintegrals}. 

 To show that $E\left[ e^{t \left( \frac{ Z_{\left(N-k+1\right)}} {b} \right)} \right]$ exists for all $t<0 $, we use use \cite [Lemma 1.7.2] {reiss2012approximate}, with $g(x)= e^{tx}$, $x \geq 0$ and $t < 0$; we have 
\begin{gather} \label{eq:71}
\begin{split}
 E\left[e^{t  \left( \frac{ Z_{\left(N-k+1\right)}} {b} \right) }\right] \leq \frac{N!}{(k-1)! (N-k)!} E \left[ e^{\frac{t Z_{i}}{b}} \right], 
\end{split}
\end{gather}
where the CDF and PDF of $Z_{i}$ are as  in  (\ref{eq:4}) and (\ref{eq:5}), respectively. We note that $E \left[ e^{\frac{t Z_{i}}{b}} \right]$ exists  for all  $t < 0$  and it  can be expressed as 
\begin{gather} \label{eq:72}
\begin{split}
E \left[ e^{\frac{t Z_{i}}{b}} \right]&=\int_{0}^{\infty}\frac{m \lambda \beta\left(P_{M}\right)^{m} z^{m-1} e^{\frac{tz}{b}} }{\left( \lambda \beta +P_{M} z \right)^{m+1}} dz\\
&=\Gamma  \left( m+1 \right) {{ e}^{-{\frac {\lambda\,\beta\,t}{2 P_{M}
}}}}{{\sl W}_{-m,\,-1/2}\left({\frac {-\lambda\,\beta\,t}{P_{M}}}\right)},  \ t<0,
\end{split}
\end{gather}
where ${\sl W}_{l,\,n}\left(\cdot \right)$ is the Whittaker function\cite{weisstein2004whittaker}. Making use of   (\ref{eq:71}) and (\ref{eq:72}), we infer that $E\left[ e^{t \left( \frac{ Z_{\left(N-k+1\right)}} {b} \right)} \right] < \infty$, for all $t<0 $. Finally, since both  $E \left[e^{t Z}\right]$ and $E\left[ e^{t \left( \frac{ Z_{\left(N-k+1\right)}} {b} \right)} \right]$  exist, based on Theorem 2 of \cite{chareka2008converse}, (\ref{eq:90})  holds.

Lemma 2 below establishes the connection between convergence in distribution and convergence of negative moments. 

\noindent \textit{Lemma 2}: If $\frac{Z_{\left(N-k+1\right)}} {b}$ converges in distribution to a random variable $Z$ whose CDF is as in (\ref{eq:11}), then  for a fixed $k$ we have 
\begin{gather}\label{eq:100}
\lim_{N\to\infty} E\left[  \left(1+\frac{Z_{\left(N-k+1\right)}}{b}\right)^{-A} \right]=   E\left[  (1+Z)^{-A} \right],  \ A >0.  
%E\left[ e^{-t \left( \frac{ Z_{\left(N-k+1\right)}} {b} \right)} \right] \to E \left[e^{-t Z}\right],
\end{gather}

\noindent \textit{Proof}: To prove (\ref{eq:100}), it is equivalent to show that  
\begin{gather}\label{eq:101}
\lim_{N\to\infty} E\left[ e^{-A \ln\left(1+\frac{Z_{\left(N-k+1\right)}}{b}\right)} \right]=   E\left[   e^{-A\ln(1+Z)} \right],  \ A >0.  
%E\left[ e^{-t \left( \frac{ Z_{\left(N-k+1\right)}} {b} \right)} \right] \to E \left[e^{-t Z}\right],
\end{gather}
Using continuous mapping theorem \cite{billingsley2008probability}, we infer  that if $\frac{Z_{\left(N-k+1\right)}} {b}$ converges in distribution to a random variable $Z$ whose CDF is as in (\ref{eq:11}); then  $\ln\left(1+\frac{Z_{\left(N-k+1\right)}}{b}\right)$ converges in distribution to $\ln(1+Z)$. 

 Theorem 2 of \cite{chareka2008converse} implies that to show that (\ref{eq:101}) holds, it suffices to show that $E\left[  (1+Z)^{-A} \right]$ and  $ E\left[  \left(1+\frac{Z_{\left(N-k+1\right)}}{b}\right)^{-A} \right]$  exist, for $A>0$. We note that  $E\left[  (1+Z)^{-A} \right]$ exists  and it can be evaluated as 
\begin{gather} \label{eq:102}
\begin{split}
 E\left[ \left( 1+ Z\right)^{-A} \right] &= \int_{0}^{\infty}   \frac{\left( 1+ z \right)^{-A}  e^{-z^{-1}} }{ z^{k+1} (k-1)!} dz\\
&=\frac{U\left(A+k; k+1;1\right) \Gamma\left(A+k\right)}{(k-1)!},  \ A>0, 
%&= \int_{0}^{\infty}   \frac{\left( 1+ b_{N} z \right)^{-A}  e^{-z^{-1}} }{ z^{k+1} (k-1)!} dz\\
\end{split}
\end{gather}
 where $\textit{U}\left( a;b;z \right)=\frac{1}{\Gamma(a)} \int_{0}^{\infty} e^{-zt} t^{a-1} (1+t)^{b-a-1}dt$, $a > 0$ is the Tricomi hypergeometric function \cite [Eq. (39)]{1576535}. The above integral is evaluated after variable transformation of  $u={z}^{-1}$ and using the definition of $\textit{U}\left( a;b;z \right)$.

To show that $E\left[  \left(1+\frac{Z_{\left(N-k+1\right)}}{b}\right)^{-A} \right]$  exists for all $A>0 $, we use \cite [Lemma 1.7.2] {reiss2012approximate}, with $g(x)= (1+x)^{-A}$, $x \geq 0$ and $A>0$; we have 
\begin{gather} \label{eq:103}
\begin{split}
 E\left[  \left(1+\frac{Z_{\left(N-k+1\right)}}{b}\right)^{-A} \right] \leq& \frac{N!}{(k-1)! (N-k)!} \\
& \times E\left[  \left(1+\frac{Z_{i}}{b}\right)^{-A} \right], 
\end{split}
\end{gather}
where the CDF and PDF of $Z_{i}$ are as in  (\ref{eq:4}) and (\ref{eq:5}), respectively. Making use of \cite[Eq. (3.197.1)] {xxx}, we note that $E\left[  \left(1+\frac{Z_{i}}{b}\right)^{-A} \right]$ exists for $ A>0$ and can be expressed as 
\begin{gather} \label{eq:104}
\begin{split}
E\left[  \left(1+\frac{Z_{i}}{b}\right)^{-A} \right]=& \frac{  m \lambda \beta b^{A}}{P_{M}} \int_{0}^{\infty}\frac{ z^{m-1} dz }{\left(b+z\right)^{A} \left( \frac{\lambda \beta}{P_{M}} + z \right)^{m+1}} \\
=& m  B \left( m,A+1 \right) \\
& \times {}_2F_1\left( A;m;A+m+1;  1- \frac{\lambda \beta}{b P_{M} } \right), 
\end{split}
\end{gather}
where ${}_2F_1\left( x;y; z;w\right)$ is the Gauss hypergeometric function \cite[Eq. (9.111)] {xxx} and $B \left( x,y \right)$ is the Beta function \cite [Eq. (8.380)]{xxx}.  

Combining  (\ref{eq:103}) and (\ref{eq:104}), we infer that $E\left[  \left(1+\frac{Z_{\left(N-k+1\right)}}{b}\right)^{-A} \right]  < \infty$ for all $A>0 $. Finally, since both $E\left[ \left( 1+ Z\right)^{-A} \right]$ and $E\left[  \left(1+\frac{Z_{\left(N-k+1\right)}}{b}\right)^{-A} \right]$   exist for $ A>0$, based on Theorem 2 of \cite{chareka2008converse}, (\ref{eq:100})  holds.

\section*{Acknowledgement} 
This publication was made possible by the NPRP award [NPRP 8-648-2-273] from the Qatar National Research Fund (a member of The Qatar Foundation). The statements made herein are solely the responsibility of the authors.
\bibliographystyle{IEEEtran}
\bibliography{yvlc}
\end{document}